\makeatletter
\declare@file@substitution{revtex4-1.cls}{revtex4-2.cls}
\makeatother
\documentclass[twocolumn]{aastex63}

\usepackage{graphicx}	

\shorttitle{Enrichment from AGB stars}
\shortauthors{R. J. Parker \& C. Schoettler}

\begin{document}

\title{Isotopic enrichment of planetary systems from Asymptotic Giant Branch stars}

\correspondingauthor{Richard Parker}
\email{R.Parker@sheffield.ac.uk}

\author[0000-0002-1474-7848]{Richard J. Parker}
\altaffiliation{Royal Society Dorothy Hodgkin Fellow}
\affiliation{Department of Physics and Astronomy, The University of Sheffield, Hicks Building, Hounsfield Road, Sheffield, S3 7RH, UK}
\author[0000-0002-2187-4570]{Christina Schoettler}
\affiliation{Astrophysics Group, Department of Physics, Imperial College London, Prince Consort Rd, London, SW7 2AZ, UK}





\begin{abstract}
Short-lived radioisotopes, in particular $^{26}$Al and  $^{60}$Fe, are thought to contribute to the internal heating of the Earth, but are significantly more abundant in the Solar System compared to the Interstellar Medium. The presence of their decay products in the oldest Solar System objects argues for their inclusion in the Sun’s protoplanetary disc almost immediately after the star formation event that formed the Sun. Various scenarios have been proposed for their delivery to the Solar System, usually involving one or more core-collapse supernovae of massive stars. An alternative scenario involves the young Sun encountering an evolved Asymptotic Giant Branch (AGB) star. AGBs were previously discounted as a viable enrichment scenario for the Solar System due to the presumed low probability of an encounter between an old, evolved star and a young pre-main sequence star. We report the discovery in Gaia data of an interloping AGB star in the star-forming region NGC2264, demonstrating that old, evolved stars can encounter young forming planetary systems. We use simulations to calculate the yields of  $^{26}$Al and  $^{60}$Fe from AGBs and their contribution to the long-term geophysical heating of a planet, and find that these are comfortably within the range previously calculated for the Solar System.


\end{abstract}



\section{Introduction}

$^{26}$Al and $^{60}$Fe are short-lived radioactive isotopes (SLRs), with half-lives of 0.7 and 2.6 Myr, respectively \citep{CastilloRogez09,Wallner15}. Their decay isotopes are often found in chondritic meteorites, some of the oldest objects in our Solar System, which suggests that these isotopes were present at the earliest epoch of planet formation around the Sun. Furthemore, $^{26}$Al, and to a lesser extent, $^{60}$Fe, are much more abundant than in the Interstellar Medium (ISM) \citep{Kita13,Cook21}, indicating that the Giant Molecular Cloud which formed the Sun was either already enhanced in these SLRs \citep{Young14}, or some mechanism delivered them to the Solar System as it was forming \citep{Ouellette10,Fatuzzo22}. 

In the scenario where the SLRs are inherited from the GMC, in order to obtain the required $^{26}$Al/$^{60}$Fe ratio, the star forming event that formed the Sun must have been sequential, with supernovae that produced $^{60}$Fe triggering subsequent generations of stars which deliver the majority of the $^{26}$Al later, via the winds of one or more Wolf-Rayet stars \citep{Gounelle12,Gounelle15}. Whilst apparently corroborated by circumstantial observational evidence that appears to show sequential triggered star formation \citep[e.g. the Upper Sco complex,][]{Preibisch99}, simulations of star formation do not predict such an efficient triggering process \citep{Dale15a}. Furthermore, it has been shown that a triggered star-forming region would dynamically merge into the region that triggered it \citep{Parker16a}, resulting in age spreads (or even age dichotomies) exceeding 10\,Myr, which are not observed \citep{Jeffries11}.  

The scenario in which SLRs are directly delivered to the Solar System usually assumes the protosolar disc is enriched by the explosion of a nearby supernova \citep{Ouellette10,Lichtenberg16b}. Whilst this mechanism merely requires the Sun to form in a single populous star cluster, massive enough to contain stars that explode as supernovae \citep{Nicholson17}, the main issue is that any supernovae are unlikely to explode until 4 Myr \citep[or much later, if stars $>$25 M$_\odot$ directly collapse to black holes rather than exploding as supernovae, e.g.][]{Limongi18}, by which time the protosolar disc will have evolved to form planets \citep[or may have been destroyed by the ionising radiation from the same massive stars that enrich the disc,][]{Nicholson19a,ConchaRamirez21}. This tension in timescales can be slightly mitigated if the low-mass disc-hosting star receiving the ejecta forms after the massive star(s), or is enriched by the winds of the massive stars \citep{Zwart19,Parker23a}. However, significant amounts of $^{60}$Fe is only produced by the supernovae of massive stars, which do not explode until $\sim$10\,Myr in the latest stellar evolution models \citep{Limongi18}.   

An alternative production channel for $^{26}$Al and $^{60}$Fe is in the cores of Asymptotic Giant Branch (AGB) stars \citep{Karakas16}. An AGB star is a post-main sequence evolutionary phase undertaken by stars with initial masses 1 -- 8 M$_\odot$ \citep{Herwig05,Ventura18}. AGB stars dredge up their interiors and drive powerful winds, making the delivery of SLRs relatively straightforward \citep{TrigoRodriguez09}. The issue with AGBs is that it was thought unlikely that an old, evolved star would have a chance encounter with the young Sun as it was forming planets \citep{Kastner94}.

However, the revolution in positional astronomy thanks to the Gaia mission has enabled researchers to accurately determine membership of star-forming regions, as well as being able to trace fast-moving `runaway' [RW] (stars moving $>$30\,km\,s$^{-1}$) and slower-moving `walkaway' [WW] stars ($>$5\,km\,s$^{-1}$). In these analyses, interloping, or visiting stars can be disentangled from the host population of the star-forming region, demonstrating that older stars could encounter younger stars, and vice versa \citep{Schoettler21}.


In this letter, we report the serendipitous discovery in Gaia DR3 of an interloping AGB star that has recently passed through a young star-forming region. We present the observational evidence in Section~2. We then model the enrichment of young stars and their protoplanetary discs using $N$-body simulations, and calculate the distribution of the yields of $^{26}$Al and $^{60}$Fe from an interloping AGB star in Section~3. We discuss caveats and conclude in Section~4.

\section{An interloping AGB star in NGC\,2264}

\begin{figure}
    \centering
    \includegraphics[width=\columnwidth]{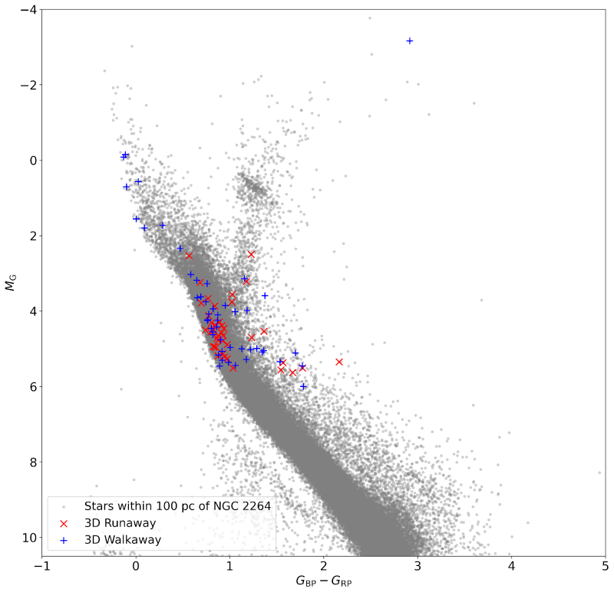}
    \caption{Gaia DR3 colour--absolute--magnitude diagram of stars within 100 pc of NGC\,2264 identifying RW ($>$30\,km\,s$^{-1}$, shown in red) and WW ($>$5\,km\,s$^{-1}$, shown in blue) stars. The WW at the top right of the giant branch (Gaia DR3 3131012157848982272) is a likely AGB star.}
    \label{fig:CMD}
\end{figure}

Whilst performing a search in Gaia Data Release 3 \citep[DR3,][]{Gaia20,Gaia22} for RW and WW stars in the vicinity of nearby star-forming regions, we discovered a WW star on the giant branch of the colour-absolute magnitude diagram (Fig.~\ref{fig:CMD}) of NGC\,2264, a young \citep[$\sim$3 Myr, e.g.][]{Dahm08}, relatively nearby \citep[$\sim$723 pc,][]{CantatGaudin20} star-forming region, postulated to have formed stars in a very dense configuration \citep{Parker22a}.

To identify interloping stars from Gaia DR3 data, we follow the method described in \citet{Schoettler22} for Gaia DR2 data but update the information on the position and velocity of the centre of NGC\, 2264 from \citet{CantatGaudin20} and \citet{Carrera19}. These central values have lower associated uncertainties compared to those used in \citet{Schoettler22}, who used a different source for consistency with previous work.

We then collect the position and velocity information for all stars within 100\,pc of this centre from the Gaia archive. Instead of parallax, we use the (photogeometric) distance estimates from \citet{BailerJones21} for all stars. We apply a rest frame centred on the above values to all data and convert the position and velocity into a cartesian reference frame. We then use a straight-line trace-back and identify any star as an interloper candidate if it can be traced back to a region within 2\,pc in the x- and y- direction of the centre \citep[on the sky search radius as derived and used in][]{Schoettler22}. We do not use the search radius definition in the z-direction (radial distance) from that analysis due to the large centre distance errors. The NGC\,2264 centre distance from \citet{CantatGaudin20} has a much smaller associated uncertainty, which translates to a search radius of 4\,pc (2\,pc as the on the sky radius + 2\,pc distance uncertainty) in the z-direction.

We apply a maximum trace-back time of 5 Myr \citep[as in][]{Schoettler22} and exclude any stars that were within our search region before that time. We then plot all the traced-back RW/WW candidates on a extinction/reddening-uncorrected colour--absolute--magnitude diagram (Fig.~\ref{fig:CMD}) and identify a star at a location inconsistent with the young age of NGC\,2264.


The WW star (Gaia DR3 3131012157848982272) was traced back to the northern region of NGC\,2264 using its proper motion and radial velocity. It has a velocity in the reference frame of NGC 2264 of $22.6 \pm 1.8$\,km\,s$^{-1}$ and likely flew through this region $\sim$3 Myr ago. Gaia DR3 identifies this star as a long-period variable (LPV), a classification which encompasses AGB stars. We use the method to identify the subclass of this potential AGB star using Gaia DR3 and 2MASS photometry developed by \citet{Lebzelter18}. Our identified WW-LPV star is located firmly within the region of O-rich low-mass AGB-stars on the ($W_{RP, BP-RP} - W_{K_s,J-K_s}$) vs. $K_s$ diagram (with the x-axis value of $\sim$0.2 mag and the y-axis value of  $\sim$10.8 mag; after calculating its $K_s$ magnitude at the LMC distance from its 2MASS $K_s$ magnitude of $\sim$1.8 mag at $\sim$760 pc). This group (O-rich and low-mass) contains stars during the early-AGB and thermally pulsing AGB (TP-AGB) phases with initial masses from $\sim$0.9 M$_\odot$ to $\sim$1.4 M$_\odot$. Its position ($W_{RP, BP-RP} - W_{K_s,J-K_s}$: $\sim$0.2 mag and $M_{K_s}$: $\sim$-7.6 mag) coincides with the evolutionary track \citep[as shown in Fig. B.1 in][]{Lebzelter18} of a star with an initial mass of $\sim$1.3\,M$_\odot$ to $\sim$1.6\,M$_\odot$.

It is unlikely that this old, evolved star formed in NGC\,2264 \citep[even the most conservative estimates place the duration of star formation at less than 10 Myr, e.g.][]{Chevance20}, and its relatively fast velocity suggests that it has moved through NGC\,2264, having formed elsewhere (we hereon in refer to this star as an `interloper'). 

\section{$N$-body simulations}

Under the assumption that some star-forming regions may host interloping AGB stars, we use $N$-body simulations and published yields \citep{Karakas14,Karakas16} to calculate the expected quantities of $^{26}$Al and $^{60}$Fe that could be accreted by a protoplanetary disc in a star-forming region with similar properties to NGC 2264. We assume the AGB star has deposited material from its winds at roughly the same time as star formation takes place, although we test this assumption by varying the size of the volume (and hence density) of AGB ejecta.

The simulations in question contain $N_\star = 1000$ stars drawn from a \citet{Maschberger13} IMF with a probability distribution of the form
\begin{equation}
p(m) \propto \left(\frac{m}{\mu}\right)^{-\alpha}\left(1 + \left(\frac{m}{\mu}\right)^{1 - \alpha}\right)^{-\beta}.
\label{maschberger_imf}
\end{equation}
Here, $\mu = 0.2$\,M$_\odot$ is the average stellar mass, $\alpha = 2.3$ is the \citet{Salpeter55} power-law exponent for higher mass stars, and $\beta = 1.4$ describes the slope of the IMF for low-mass objects \citep*[which also deviates from the log-normal form;][]{Bastian10}. We randomly sample this distribution in the mass range 0.1 -- 50\,M$_\odot$.

The resultant total mass for the star-forming regions are of order $\sim$500\,M$_\odot$ (there is some variation due to the stoachastic nature of sampling the stellar IMF), which lies towards the lower end of the observed mass function for star-forming regions \citep{Lada03} and is similar to the mass of NGC\,2264, the region in which we have observed an interloping AGB star. Such regions are more common than their higher-mass counterparts (e.g.\,\,Westerlund~1, R136), but rarer than low-mass star-forming regions (e.g.\,\,Taurus, $\rho$~Oph). 

The stars are all single (i.e. we do not include primordial binaries, and this simplification is unlikely to affect the numbers of stars that would encounter the interloping AGB star). 

The simulated star-forming regions have initial radii of 1 pc, which results in a median stellar density of 1000\,M$_\odot$\,pc$^{-3}$. Pre-stellar cores, and pre-main sequence stars, are observed in spatially and kinematically substructured distributions \citep{Andre14,Hacar13,Cartwright04} and for this reason we set up our simulations as self-similar fractal distributions \citep{Goodwin04a}, where the stars' positions and velocities are correlated according to a fractal dimension $D = 2.0$. Similarly, the star-forming regions are initially set up with subvirial velocities where the virial ratio $\alpha_{\rm vir} = T/|\Omega|$, and $T$ and $|\Omega|$ are the total kinetic energy and total potential energy of the stars, respectively.  In this definition, $\alpha_{\rm vir} = 0.5$ means the system is in virial equilibrium. We adopt $\alpha_{\rm vir} = 0.3$ to mimic the initial velocity distributions in observed \citep[e.g.][]{Foster15} and simulated \citep[e.g.][]{Bate12} star-forming regions.

\begin{figure}
    \centering
    \includegraphics[width=\columnwidth]{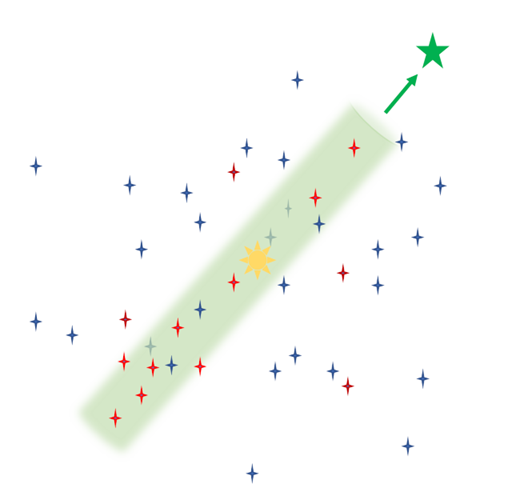}
    \caption{Sketch of the assumed geometry of an interloping AGB star in a star-forming region. The AGB star (green) has traversed the star-forming region, depositing a cylindrical shaped ejecta. Some stars (red) are enriched and then leave the ejecta region, whereas others remain. Blue stars are those that do not capture any of the ejecta (including fore- and background stars that reside outside of the ejecta). The hypothetical young Sun is shown by the yellow star. 
}
    \label{fig:AGB_sketch}
\end{figure}

We assume AGB stars that are identified as runaway and walkaway stars will have a similar velocity distribution to stars ejected from star-forming regions, and so we select the median velocity from the distribution in \citet{Schoettler19}. We assume the AGB star traverses the star-forming region almost immediately after star formation (such that the protoplanetary discs are yet to form planets). We note that this occurred for the AGB star that passed through NGC\,2264; it interloped through the region 3\,Myr ago, which is roughly the age of the stars in the region.

The ejecta from the wind of the AGB star is modelled as a cylinder with length four times the half-mass radius of the star-forming region, $l_{\rm ejecta} = 4r_{1/2}$ (note that the half-mass radius expands as the star-forming region evolves, meaning that the cylinder of AGB ejecta also increases in length, reducing the density of the ejecta). We experimented with varying the radii of the cylinder of ejecta, $r_{\rm ejecta}$; smaller radii (e.g. $r_{\rm ejecta} = 0.1$\,pc) lead to high-level enrichment of a few stars, whereas larger radii (0.5 -- 1\,pc) lead to enrichment of more stars, but at lower values.

A cartoon of the AGB ejecta geometry is shown in Fig.~\ref{fig:AGB_sketch}. The AGB star (green) has traversed the star-forming region, depositing a cylindrical shaped ejecta. Some stars (shown in red) are enriched and then leave the ejecta region, whereas others remain. Blue stars are those that do not capture any of the ejecta (including fore- and background stars that reside outside of the ejecta). The hypothetical young Sun is shown by the yellow star.

We determine whether a star crosses through the AGB ejecta, and if it does we calculate the distance it travels, $d_{\rm trav}$. We use this, and the radius of the disc around the star, $r_{\rm disc}$, to determine the amount of AGB ejecta swept up from the cylinder:
\begin{equation}
\eta_{\rm ejecta} = \frac{r_{\rm disc}^2d_{\rm trav}}{l_{\rm ejecta}r_{\rm ejecta}^2}.
\end{equation}
This equation assumes that none of the AGB ejecta is deflected by the disc \citep[i.e.\,\,the relative velocities are low, unlike when a disc encounters a supernova blast wave,][]{Ouellette07}. Furthermore, this represents the maximum possible amount of material swept up, as we do not account for the inclination angle of the disc \citep[although the average angle is likely to be in the region of 60$^\circ$,][which would only reduce the enrichment by a factor of two]{Lichtenberg16b}.

We then determine the mass of $^{26}$Al, $m_{ ^{26}{\rm Al}}$, swept up by the protostellar disc by dividing the total $^{26}$Al yield from the AGB star, $m_{ ^{26}{\rm Al, AGB}}$, by the time spent in the AGB phase $t_{\rm AGB}$ and then multiplying this by the time taken for the star to travel $d_{\rm trav}$, which is $\Delta t$:
\begin{equation}
m_{ ^{26}{\rm Al}} =  \eta_{\rm ejecta} \frac{m_{ ^{26}{\rm Al, AGB}}}{t_{\rm AGB}}\Delta t.
\end{equation}
We perform a similar calculation to determine the mass of $^{60}$Fe swept up by the disc, $m_{ ^{60}{\rm Fe}}$:
\begin{equation}
m_{ ^{60}{\rm Fe}} =  \eta_{\rm ejecta} \frac{m_{ ^{60}{\rm Fe, AGB}}}{t_{\rm AGB}}\Delta t, 
\end{equation}
where $m_{ ^{60}{\rm Fe, AGB}}$ is the amount of $^{60}$Fe produced by the AGB star, and the time variables are as above. 

In order to calculated the amount of $^{26}$Al and $^{60}$Fe relative to the stable isotopes in the protoplanetary discs ($^{27}$Al and $^{56}$Fe), we make some assumptions about the masses and sizes of the protoplanetary discs. 


For each planet-hosting star (masses between $0.1 < M_\star/{\rm M_\odot} < 3$\,M$_\odot$), we assign it a disc of mass
 \begin{equation}
   M_{\rm disc} = 0.1\,M_\star,
 \end{equation}
 and a radius $r_{\rm disc} = 400$\,au, commensurate with the observed discs in nearby star-forming regions \citep[e.g.][]{Andrews13,Ansdell16,Barenfeld17,Tazzari17,Eisner18,Cieza19}. We assume that the discs do not lose mass, nor do their radii evolve inwards due to external photoevaporation from massive stars, or outwards due to viscous spreading. We assume a gas-to-dust ratio of 100:1 such that the mass of solids in the disc is
 \begin{equation}
m_{\rm dust} = 0.01\,M_{\rm disc}.
 \end{equation}
The amount of dust that is $^{27}$Al is given by 
 \begin{equation}
m_{^{27}{\rm Al}} = 8500\times10^{-6}m_{\rm dust},
 \end{equation}
 and the amount of dust that is $^{56}$Fe is given by
\begin{equation}
m_{^{56}{\rm Fe}} = 1828\times10^{-4}m_{\rm dust},
\end{equation}
\citep{Lodders03}. We then use the mass of SLR swept up by the disc to calculate the yields of $^{26}$Al and $^{60}$Fe thus:
 \begin{equation}
Z_{\rm Al} =   \frac{m_{^{26}{\rm Al}}}{m_{^{27}{\rm Al}}},
 \end{equation}
  \begin{equation}
Z_{\rm Fe} =   \frac{m_{^{60}{\rm Fe}}}{m_{^{56}{\rm Fe}}}.
   \end{equation}





\begin{figure}
    \centering
    \includegraphics[width=\columnwidth]{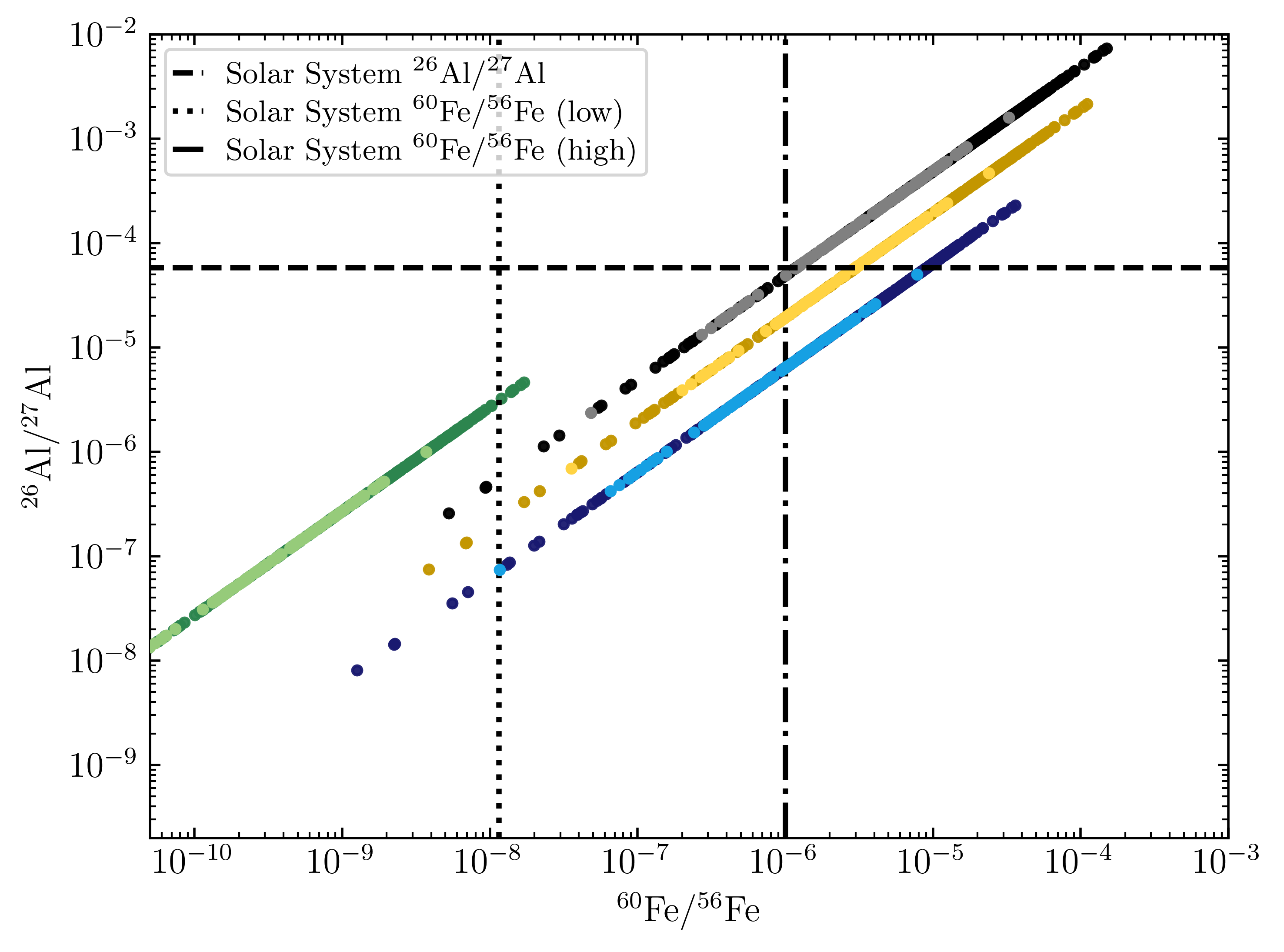}
    \caption{Abundance ratios of $^{26}$Al against abundance ratios of $^{60}$Fe in $N$-body simulations where the stars have encountered ejecta from an AGB star. The coloured points indicate different AGB progenitor masses; black are 7\,M$_\odot$ stars, yellow are 6\,M$_\odot$, blue are 5\,M$_\odot$ and green are 3\,M$_\odot$. The darker points show the values for all stars, whereas the lighter points show the values for Sun-like (0.5 -- 1.5\,M$_\odot$) stars. The horizontal dashed line indicates the initial  $^{26}{\rm Al}/^{27}{\rm Al}$ value in the Solar System inferred from CAI inclusions in chondritic meteorites \citep{Thrane06}. The vertical dot-dashed line \citep{Mishra16}, and dotted line \citep{Tang12}, indicate the possible range of the initial  $^{60}{\rm Fe}/^{56}{\rm Fe}$ value in the Solar System (this measurement is more uncertain).}
    \label{fig:al26_fe60}
\end{figure}

  In Fig.~\ref{fig:al26_fe60} we show the $^{26}$Al/$^{27}$Al ratio as a function of the $^{60}$Fe/$^{56}$Fe ratio, with the initial $^{26}$Al/$^{27}$Al ratio measured in the Solar System shown by the horizontal dashed line \citep{Thrane06}.  The measurement of the abundance of $^{60}$Fe in the early Solar System is more controversial, with estimates varying by several orders of magnitude, from $^{60}$Fe/$^{56}$Fe $\sim 10^{-8} - 10^{-6}$ \citep{Tang12,Mishra16,Cook21,Kodolanyi22}. The maximum range of values for the initial Solar System  $^{60}$Fe/$^{56}$Fe ratio is shown by the vertical dot-dashed line \citep{Mishra16}, and vertical dotted line \citep{Tang12}. 

We plot the $^{26}$Al/$^{27}$Al ratio and the $^{60}$Fe/$^{56}$Fe ratio for different AGB progenitor masses; the black points are for 7\,M$_\odot$, the yellow points for 6\,M$_\odot$, the blue points for 5\,M$_\odot$ and the green points are for 3\,M$_\odot$ progenitor masses. All other parameters (density of AGB ejecta, radius of accreting protoplanetary disc, etc.) are kept constant. The lighter coloured points indicate stars of roughly Solar-mass (0.5--1.5 M$_\odot$). Fig.~\ref{fig:al26_fe60} indicates that a reasonably high initial AGB progenitor mass (6--7\,M$_\odot$) is required to provide Solar System-like abundances, although these values can also change if we increase the stellar density in the star-forming region, or the radius of the accreting protoplanetary disc. Note that the AGB star we have found interloping through NGC\,2264 has a lower progenitor mass ($\sim1.5$\,M$_\odot$), and such stars are more common than the 6--7\,M$_\odot$ star needed to enrich the Solar System. However, we merely wish to demonstrate here that AGB stars can interlope through star-forming regions, and a higher mass AGB star can provide the enrichment found in the Solar System.

At this stage, we do not account for the radioactive decay of the SLRs. This is because we assume the AGB desposits material at the instant of star formation in our simulations, and due to the relatively high stellar density, most ($>$90\%) of the AGB ejecta is swept up in the first 0.5--1\,Myr.

The presence of short-lived radioisotopes provides an additional heat source in the interiors of planets, especially if the SLRs are incorporated early in the formation of the planetary system, i.e. before differentiation within the individual bodies has occurred. The contribution from $^{26}$Al and $^{60}$Fe dominated the radiogenic heat budget of the early Earth  \citep{McDonough20}, with the amount of heating calculated at $Q(t) = 3.5 \times 10^{-7}$\,W\,kg$^{-1}$, assuming $^{26}{\rm Al}/^{27}{\rm Al} = 5.85 \times 10^{-5}$ \citep{Thrane06} and $^{60}{\rm Fe}/^{56}{\rm Fe} = 1 \times 10^{-6}$ \citep{Mishra16}.

For each star in our simulations, we use the abundance of $^{26}$Al (defined by the ratio $^{26}$Al/$^{27}$Al) and the abundance of $^{60}$Fe (defined by the ratio $^{60}$Fe/$^{56}$Fe) to calculate the geophysical heating \citep{Moskovitz11} at a given time after the AGB star has deposited the SLRs in the star-forming region (here, we do account for the decay of the SLRs)  using 
\begin{equation}
  Q(t) = f_{\rm Al,CI}Z_{\rm Al}\frac{E_{\rm Al}}{\tau_{\rm Al}}e^{-t/\tau_{\rm Al}} + f_{\rm Fe,CI}Z_{\rm Fe}\frac{E_{\rm Fe}}{\tau_{\rm Fe}}e^{-t/\tau_{\rm Fe}},
  \label{ss_heating}
  \end{equation}
where $f_{\rm Al,CI}$ is the fraction of Al in chondrites \citep{Lodders03}, $E_{\rm Al} = 3.12$\,MeV is the decay energy of $^{26}$Al, $\tau_{\rm Al} = 0.717$\,Myr is the radioactive half-live of $^{26}$Al  \citep{CastilloRogez09}. Similarly, $f_{\rm Fe,CI}$ is the fraction of Fe in chondrites \citep{Lodders03}, $E_{\rm Fe} = 2.712$\,MeV is the decay energy of $^{60}$Fe \citep{CastilloRogez09} and  $\tau_{\rm Fe} = 2.6$\,Myr is the half-life of $^{60}$Fe \citep{Wallner15}. The initial Solar system heating value is calculated from these values to be $Q_{\rm SS} = 3.4 \times 10^{-7}$\,W\,kg$^{-1}$.



\begin{figure}
    \centering
    \includegraphics[width=\columnwidth]{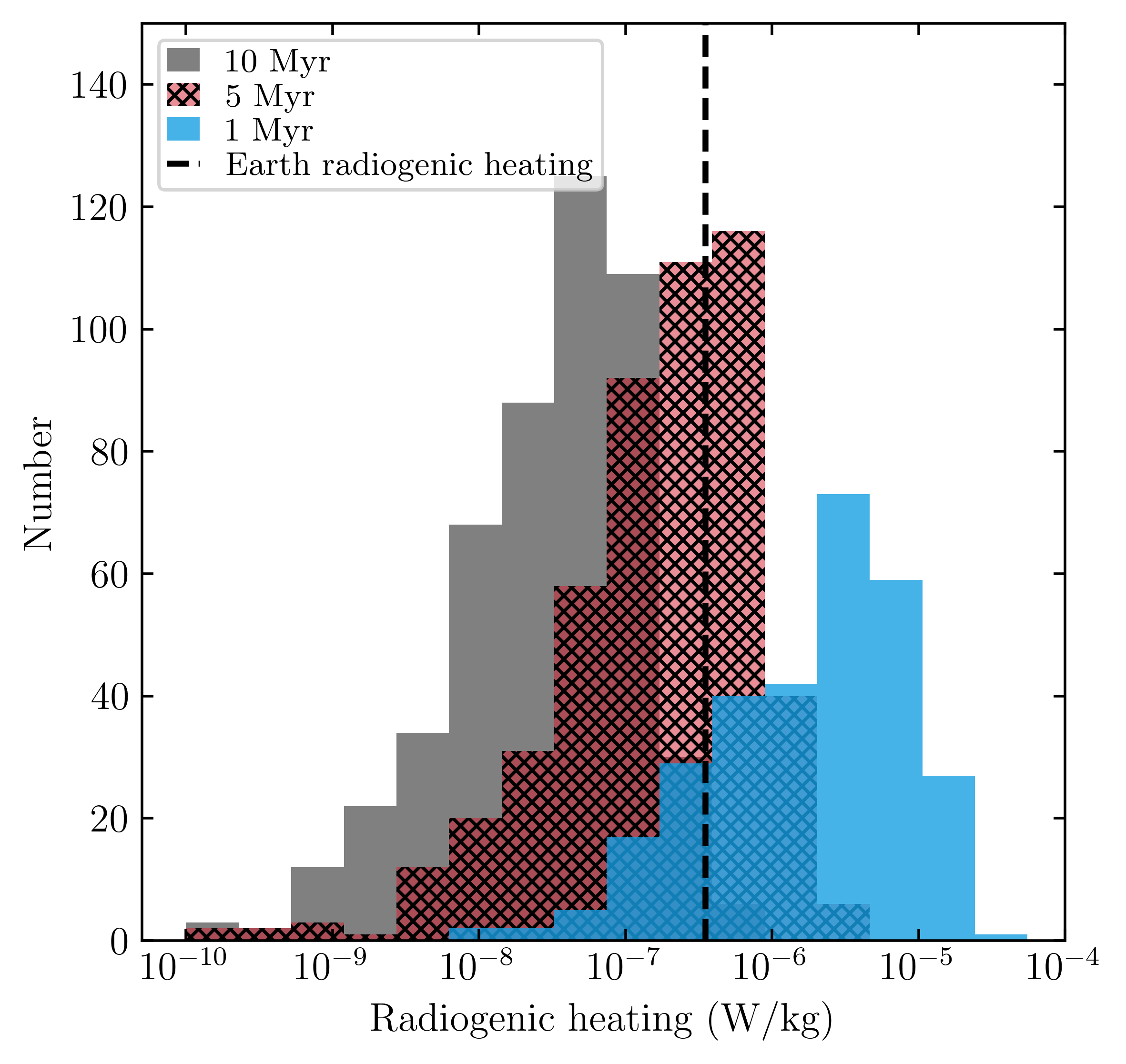}
    \caption{The long-term internal geophysical heating of planessimals, calculated form the relative abundance ratios of $^{26}$Al and $^{60}$Fe in our $N$-body simulations where the stars have encountered ejecta from an AGB star. The AGB ejecta is assumed to have been deposited immediately after the onset of star formation, and the heating, $Q(t)$ is calculated at 1, 5 and 10 Myr (blue,  hatched red, and grey histograms, respectively). The vertical dashed line indicates the likely heating value for the early Solar System immediately after the inclusion of short-lived radioisotopes in the Sun’s protoplanetary disc \citep{Moskovitz11}. The highest degree of internal heating occurs at earlier times, before the majority of the $^{26}$Al and $^{60}$Fe has decayed. }
    \label{fig:heating_hist}
\end{figure}

We calculate the geophysical heating at various times after star formation.  We assume that this is the same length of time after the production of AGB ejecta, and the resultant distributions are shown in Fig.~\ref{fig:heating_hist}. The different histograms correspond to the heating after 1 (blue), 5 (red), and 10 Myr (grey), respectively. The vertical dotted line indicates the heating value, $Q(t)$, calculated for the early Solar System \citep{Moskovitz11}. The early heating of protoplanetary discs enriched by AGB ejecta is consistent with the value calculated for the Solar System, and we find that several combinations of our initial conditions (simulation density, AGB progenitor mass, protoplanetary disc radius, density of AGB ejecta) are within this range.


\section{Discussion and conclusions}
\label{sec:conclusions}

The main caveat to our results is that the proportion of $^{60}$Fe in comparison to $^{26}$Al is towards the high end of the range of values for the measured ratio in the Solar System. This could be alleviated if the enriching AGB star has a lower progenitor mass (in some cases the ratio of $^{26}$Al/$^{60}$Fe is then higher, e.g. the green points in Fig. 3, which are the yields from a 3 M$_\odot$ progenitor star), and there is an additional source of $^{26}$Al, e.g. from the wind(s) of massive stars in the star-forming region. However, the latter scenario would then require the Sun to form in a star-forming region also containing massive stars, which somewhat limits the advantages of enrichment from AGB stars. This would also return us to the arguments against the direct enrichment of the Solar System by massive stars (only a small fraction of star-forming regions contain massive stars, and those massive stars’ FUV and EUV radiation fields could evaporate planet-forming discs). 

Our results depend slightly on the volume density of the AGB ejecta; the lower the density of the ejecta, the lower the enrichment. In our default simulations, the ejecta is dispersed in a cylinder of 0.1pc, but Solar System levels of enrichment can occur when the cylinder has a larger radius. Furthermore, the most likely AGB progenitor mass is between 5 -- 8\,M$_\odot$, but we note that a lower progenitor mass is possible if the star-forming region is initially very dense \citep[as is the case for NGC\,2264,][]{Schoettler22}. We do not know the initial density of the star-forming region in which the Solar System formed, but based on the frequency of dynamical interactions that would truncate the Sun’s protosolar disc and disrupt the early Solar System, the initial stellar density could be slightly higher than those we adopt in our simulations \citep[$>10^4$\,M$_\odot$\,pc$^{-3}$,][]{Pfalzner20}. 

We also ran test simulations where we lowered the stellar density to 100\,M$_\odot$\,pc$^{-3}$. In this scenario, a similar amount of material is swept up, but the enrichment occurs later, after around 5\,Myr, once the star-forming region has attained its maximum density. We would therefore expect some of the $^{26}$Al and $^{60}$Fe to have already decayed, unless the AGB star deposited the material several Myr after star formation. 

If we instead reduce the number of stars (but keep the stellar density constant at 1000\,M$_\odot$\,pc$^{-3}$), then for a single star-forming region fewer stars are enriched overall, but the fraction of stars that are enriched is very similar to the simulations with higher $N_\star$ (because the stellar densities are similar).

The size of the protoplanetary disc in our simulations is fixed at 400 au, commensurate with the disc sizes measured in nearby star-forming regions. As expected, reducing the initial disc radii reduces the cross section for enrichment, but we note that viscous expansion of smaller discs means that our adopted disc radii could be conservative estimates after several Myr of evolution \citep{ConchaRamirez19}. 

In summary, we have shown that evolved stars can encounter forming planetary systems, and the yield of $^{26}$Al and $^{60}$Fe from AGB stars can account for the enrichment and subsequent geophysical heating in the Solar system. \\


\section*{Acknowledgements} 
We thank the anonymous referee for a prompt and helpful report. RJP acknowledges support from the Royal Society in the form of a Dorothy Hodgkin Fellowship. This work has made use of data from the European Space Agency (ESA) mission Gaia (https://www.cosmos.esa.int/gaia), processed by the Gaia Data Processing and Analysis Consortium (DPAC, https://www.cosmos.esa.int/web/gaia/dpac/consortium). Funding for the DPAC has been provided by national institutions, in particular the institutions participating in the Gaia Multilateral Agreement.  This project has received  funding from the European Research Council (ERC) under the European Union’s Horizon 2020 research and innovation programme (Grant agreement no. 853022, PEVAP).







\bibliographystyle{aasjournal}
\bibliography{general_ref} 




\end{document}